\documentclass[aps,preprint,prd,showpacs,nofootinbib]{revtex4}
\usepackage{amsmath}
\usepackage{graphicx}
\usepackage{dcolumn}
\usepackage{bm}
\usepackage{amssymb}
\usepackage{latexsym}
\usepackage{color}

\def\be{\begin{equation}}
\def\ee{\end{equation}}
\def\ba{\begin{eqnarray}}
\def\ea{\end{eqnarray}}

\begin{document}

\title{ AdS cycles in eternally inflating background }

\author{Zhi-Guo Liu$^{1,}$\footnote{Email: liuzhiguo08@mails.ucas.ac.cn}}
\author{Yun-Song Piao$^{1,2}$\footnote{Email: yspiao@ucas.ac.cn}}

\affiliation{$^1$ School of Physics, University of Chinese Academy
of Sciences, Beijing 100049, China}

\affiliation{$^2$ State Key Laboratory of Theoretical Physics, Institute of Theoretical Physics, \\
Chinese Academy of Sciences, Beijing 100190, China}

\begin{abstract}

In the eternally inflating background, the bubbles with AdS vacua
will crunch. However, this crunch might be followed by a bounce.
It is generally thought that the bubble universe may be cyclic,
which will go through a sequence of AdS crunches, until the field
inside bubble finally lands at a dS minimum. However, we show that
due to the amplification of field fluctuation, the bubble universe
going through AdS cycles will inevitably fragment within two or
three cycles. We discuss its implication to the eternal inflation
scenario.

\end{abstract}

\maketitle

\section{Introduction}

During the eternal inflation \cite{V1}, an infinite number of
bubble universes with different dS and AdS vacua will spawn. The
bubble universe with a dS minimum will arrive at a dS regime
asymptotically, while the universe with AdS minimum will collapse
rapidly. Inside an observational bubble universe, a phase of the
slow-roll inflation and reheating is required, which will set the
initial conditions of the ``big bang" evolution, i.e. a
homogeneous hot universe with the scale invariant primordial
perturbation.

The slow roll inflation should occur in a high energy scale, which
is required to insure that the amplitude of primordial
perturbation is consistent with the observations and the reheating
temperature is suitable for a hot big bang evolution after
inflation. However, if the scale of the eternally inflating
background is very low, the spawning of observational universe
will be island-like, which is exponentially unfavored, since it
requires a large upward tunneling, e.g.
\cite{GV},\cite{DV},\cite{island},\cite{Aguirre:2005xs},\cite{Lee:2013mza},
and also
\cite{Rubakov:2013kaa},\cite{Liu:2013xt},\cite{Hwang:2010gc} for
an alternative study.

Recently, it has been argued that the nonsingular bounce in the
eternally inflating background might significantly alter this
result \cite{Garriga:2012bc}, and also
\cite{Piao:2004me},\cite{Piao:2009ku},\cite{Johnson:2011aa},\cite{Garriga:2013cix},\cite{Gupt:2013poa}.
In this scenario, the crunch of AdS bubble will be followed by a
bounce, which makes the field inside bubble be able to finally
land at a dS minimum, thus we actually may have a transition from
AdS to dS
\cite{Piao:2004me},\cite{Garriga:2013cix},\cite{Gupt:2013poa}. The
introduction of AdS bounce insures that the timelike geodesics in
the eternally inflating spacetime dose not end at the big crunch
singularity inside AdS bubbles, which may make the eternally
inflating spacetime allow for a well-defined watcher measure
\cite{Garriga:2012bc}, in which what is counted is the
observations made by a single observer at a timelike geodesic, and
also close related Refs.\cite{GV},\cite{Vanchurin:2006xp}. In
addition, AdS bounce also brings an efficient route to the
slow-roll inflation. The bounce inflation may explain not only the
power deficit on large angular scales, but also a large dipole
power asymmetry in CMB
\cite{Piao:2003zm},\cite{Liu:2013kea},\cite{Biswas:2013dry}
observed by the Planck collaboration, see also
\cite{Xia:2014tda},\cite{Qiu:2014nla},\cite{Mielczarek:2008pf}.

%instead of counting observations made by all observers within a
%spacetime region defined by a geometric cutoff, we include only

However, if the field inside bubble lands at a AdS minimum of its
effective potential, the bubble universe will collapse and bounce
again. Thus the bubble universe may go through a sequence of AdS
crunches, during which it is cyclic, until the field inside bubble
finally arrives at a dS minimum.
%the corresponding field might go through a series of AdS minimum,
%during which the bubble universe is cyclic, until it finally
%arrives at a dS minimum.
Here, for convenience we call such a cyclic evolution as AdS
cycles. Recently, the cosmological cyclic scenario, in which the
universe goes through the periodic sequence of contraction and
expansion \cite{Tolman}, has been rewaked
\cite{STS},\cite{Barrow:1995cfa},\cite{Kanekar:2001qd}, which has
leaded to significant insights for the origin of the observable
universe. In Ref.\cite{Piao:2004me}, it is showed that during
different cycles in cyclic universe, the universe may be in
different vacua of a landscape, in which the inflation after
bounce is responsible for the emergence of observational universe.
In Ref.\cite{Sahni:2012er}, it is showed that the inflation after
bounce causes the cosmological hysteresis, which will lead to the
increase of the amplitude of cycles. The cyclic or oscillating
universe model also have been studied in Refs.
\cite{Piao:2004hr},\cite{Xiong:2008ic},\cite{Duhe:2013taa},\cite{Bamba:2012gq}.

It is generally thought that the background of cyclic universe is
homogeneous cycle by cycle all along. However, when the
perturbation is considered, the case will be altered
\cite{Piao:2009ku},\cite{Zhang:2010bb}, and also
\cite{Graham:2011nb},\cite{Lehners:2008qe}. The amplitude of
curvature perturbation on large scale is increasing in the
contracting phase, while it is almost constant in the expanding
phase. Thus the net result of one cycle is that the amplitude of
perturbation is amplified. This amplification of perturbation will
be multiplied cycle by cycle, which will eventually lead that the
homogeneity of background is destroyed.

Thus it is hardly possible that the AdS cycles of the bubble
universe will continue all along until the field inside the bubble
finally arrives at a dS minimum.
%In Refs.\cite{Garriga:2012bc},\cite{Piao:2009ku}, it is argued
%that when the background becomes highly inhomogeneous, different
%regions will correspond to different universe separated by domain
%walls.
%Instead the universe will potentially be separated into some
%segments independent of one another, each of which will evolve
%independently, as has been argued.
How the amplification of perturbation affects the evolution of
bubble universe in eternally inflating background is still
interesting to study in details.

Recently, the amplification of field perturbation, which is
induced by the self-interaction of field, has been studied in
Ref.\cite{Garriga:2013cix}.
%, in which it is showed that the
%amplification is efficient only if the field is relatively
%strongly coupled.
Here, we will concentrate on the amplification
of field perturbation induced by the amplification of the
curvature perturbation, which will generally arise when the bubble
universe goes through AdS cycles.

%Here, we will review the evolution of field inside the bubble
%universe during AdS cycles in Sec.II, and will investigate the
%evolution of field perturbation in Sec.III. In Sec.IV,
We will show that due to the amplification of field fluctuation,
the bubble universe going through AdS cycles will inevitably
fragment within two or three cycles, and a number of new ``bubble"
universes will come into being from these fragments.
%We argue that the
%proliferation of bubble universe may be universal for any
%landscape, and discuss its implication for the eternal inflation
%scenario.
%The emergence
%probability of new ``bubbles" is not expotentially suppressed.
In the eternally inflating scenario, compared to the nucleation of
bubbles in dS background, the proliferation of bubble universe
during AdS cycles is obviously more rapid, which will help the
eternal inflation to more rapidly populate the whole landscape,
wherever initially it happens.

\section{Review of AdS cycles in the landscape}

%\subsection{AdS cycles}

We will firstly review the main result of the classical evolution
of field during the AdS bounce, see also
\cite{Piao:2004me},\cite{Garriga:2013cix},\cite{Gupt:2013poa}.

%When the field is one among dS minima of its effective potential,
%the universe is in an eternal inflationary phase.
In eternally inflating background, the AdS bubble universe
nucleated will inevitably collapse. The initial conditions of the
bubble universe is set by the instanton. We follow
\cite{Garriga:2013cix}. Initially the bubble universe is in a
phase dominated by the curvature $\rho_{Cur}\sim 1/a^2$, in which
$a\sim t$ and $\phi$ is overdamped and approximately constant.
When $t\sim 1/M_{\phi}$, the field $\phi$ begins to oscillate
around its minimum with $\rho_{Mat}=\rho_{Mat0}/a^3$, which is
equivalent to the matter with the state equation $w\simeq 0$.
Here, \be \rho_{Mat0}<\rho_{Cur} \label{MC}\ee is assumed. This
implies that the evolution is approximately that of a AdS universe
with the negative curvature, \be a\simeq {1\over H_{\Lambda}}Sin
\left(H_{\Lambda}t\right), \ee where
$H_{\Lambda}=\sqrt{|\Lambda_*|\over 3}$ and $\Lambda_*$ is the
depth of AdS minimum. The expansion ends at $t= {\pi\over
2H_{\Lambda}}$ and is followed by the contraction dominated by
$\rho_{Cur}$, i.e. the curvature phase. Before the bounce, the
bubble universe may be still in a curvature phase, or a kinetic
phase dominated by ${\dot\phi}^2$.

%the field $\phi$ will oscillate around its minimum with
%$\rho_{Osi}\sim 1/a^3$, which is equivalent to the matter with the
%state equation $w\simeq 0$. Hereafter, we define $\rho_{Osi}$ as
%$\rho_{Mat}$.

When the bounce scale is larger than the potential barrier, the
field will be able to stride over the barrier. After the bounce,
${\dot\phi}^2\sim 1/a^6$ will be rapidly diluted and the field
will eventually land at a different place of its effective
potential. Thus the field $\phi$ will walk certain distance during
the AdS bounce, i.e. one single AdS cycle. Here, ``land" means
that the effective potential of field begins to become dominated
again. We define this displacement of $\phi$ as $\Delta\phi$, and
have \cite{Piao:2004me},\cite{Sahni:2012er} \ba {\Delta\phi} &
\simeq & {M_P\over \sqrt{6}}\ln({H_{B}^4\over H^2_{{Kin}}H^2_{{
Land}}})
%\nonumber\\ & = & {1\over
%\sqrt{6}}M_P\ln({\rho_{B}^2\over\rho_{ a_{Kin}}\,\,\rho_{a_{
%Land}}})
, \label{De}\ea where `Kin' defines the beginning of the kinetic
phase, which is the end of the matter contracting phase, and
$H_B^2=\rho_B/3$ is defined. The result is consistent with that of
Ref.\cite{Garriga:2013cix}. We may assume $H_{{Kin}}\sim H_{{
Land}}$, both which are generally smaller than $H_B$. Thus we have
\be \Delta\phi \gtrsim M_P.
\label{De}\ee %This implies that the field will walk to a distant
%place of the landscape.
During this period the number of hills and
valleys the field flies over is determined by the detail of the
landscape.

When the field lands at the dS minimum, the bubble universe will
be in a dS state. However, if the field lands at a AdS minimum of
its effective potential, the universe will collapse again, which
will be followed by the AdS bounce again. The bubble universe may
go through a sequence of AdS crunches, during which it is cyclic,
called as AdS cycles, until the field finally lands at a dS
minimum. In Appendix, a ``toy" model of AdS cycles is introduced.

%However, if the perturbation of field is considered, the result is
%actually not so.

%the field $\phi$ will oscillate around its minimum with
%$\rho_{Osi}\sim 1/a^3$, which is equivalent to the matter with the
%state equation $w\simeq 0$. Hereafter, we define $\rho_{Osi}$ as
%$\rho_{Mat}$.

\begin{figure}[htbp]
\includegraphics[scale=2,width=7.0cm]{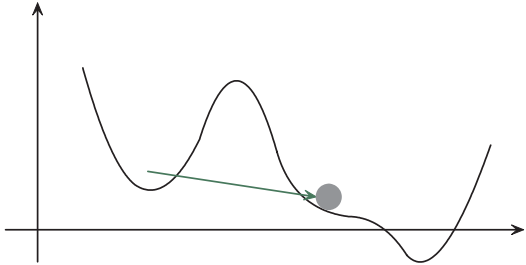}
\caption{The sketch of an effective potential.
}\label{fig:potential3}
\end{figure}

During the contraction of AdS cycles, the amplification of
$\rho_{Mat}\sim 1/a^3$ is faster than $\rho_{Cur}$. When
$\rho_{Mat}>\rho_{Cur}$, the bubble universe may be in a phase
dominated by $\rho_{Mat}$, i.e. the matter contraction. We will
see that the matter contracting phase is significant to rapidly
amplify the perturbation. However, when $\rho_{Mat} \sim
\rho_{Mat0}$, we have $H\sim M_{\phi}$, which implies the
oscillating phase ends. Thus it seems hardly possible that we have
the matter contracting phase in the $1^{th}$ cycle
\cite{Garriga:2013cix}.

However, after the $1^{th}$ cycle the initial conditions of the
field evolution will be set not by the instanton, but by the
details of previous cycle.
%the case will be altered. This may
%occur in the $2^{th}$ cycle, in which the field lands again at a
%AdS minimum of its effective potential.
In this case, the oscillating energy $\rho_{Mat}$ of field will
may be large, thus we may have the matter contracting phase in the
cycles after the $1^{th}$ cycle.

However, we may also have the matter contracting phase in the
$1^{th}$ cycle when relaxing the assumption (\ref{MC}). In
Fig.\ref{fig:potential3}, we plot a potential, for which after the
bubble nucleates, the field may be placed in a plain of its
potential. In this case, the bubble universe may has a short
inflationary phase, which will rapidly dilute $\rho_{Cur}$.
Hereafter, the field rolls toward AdS minimum, and begins to
oscillate around it. The evolution will be that of a AdS universe
with the oscillating energy $\rho_{Mat}$.

%\be a^{3/2}\simeq {\sqrt{\rho_{Mat}/3}\over H_{\Lambda}}Sin
%\left({3\over 2}H_{\Lambda}t\right). \ee Thus at $t= {\pi\over
%3H_{\Lambda}}$, the expansion ends, and the matter contracting
%phase begins.

It is generally thought that the AdS cycles will continue all the
time until the bubble universe finally ``transits" to a dS state.
However, if the perturbation of field is considered, the scenario
will be altered.

%We will show that due to the increase of field perturbation, in
%the second cycle the bubble universe will inevitably fragment,
%which results in the emergence of some new ``bubble" universes
%with different vacua.

%Thus the numerical outcomes are consistent with the analytical
%results in II.A,

\section{The proliferation of bubble universe }

\subsection{Amplification of perturbations through cycles}

We will investigate the evolution of the scalar field perturbation
through cycles. Here, the mode spectrum of the perturbation may
involve some supercurvature modes, see
e.g.\cite{Garriga:1991tb},\cite{Garriga:1996pg}. However, for
simplicity, we will only concentrate on modes whose initial
wavelength is smaller than the radius of curvature, and also on
almost flat case. In this case, the calculation of the
perturbation may be similar to that in flat case.

%negligible when the field is far away from the curvature phase.
%We will work in longitudinal gauge. The
%perturbation equations are \ba \delta\ddot\varphi_k & + &
%3\left({{\dot a}\over
%a}\right)\delta\dot\varphi_k+\left(\frac{k^2}{a^2}+
%{\partial^2 V\over \partial \varphi^2}\right)\delta\varphi_k\nonumber\\ & = & 4\dot\Phi_k\dot\varphi-2\Phi_k {\partial V\over \partial \varphi},\label{u1}\\
%-3H\dot\Phi &+& \left(\frac{\nabla^2}{a^2}-3H^2\right)\Phi \nonumber\\
%& =& {1\over
%2M_P^2}\left(\dot\varphi\delta\dot\varphi-\dot\varphi^2\Phi+{\partial
%V\over \partial \varphi}\delta\varphi\right),\label{u2}
%\\ \dot\Phi+H\Phi & = & {1\over
%2M_P^2}\dot\varphi\delta\varphi, \label{u3}
%\ea where $\Phi$ is the metric perturbation.

In the perturbation equation of scalar field $\phi$, the terms
${\dot \phi}{\dot \Phi}$ and $\Phi {\partial V\over \partial
\phi}$, coupling to the metric perturbation $\Phi$, should be not
negligible during the contraction, since both $\dot \phi$ and
$\Phi$ are rapidly increasing. The equations of perturbations are
a set of coupling equations between $\Phi$ and $\delta\phi$, thus
solving the equation of $\delta\phi$ has to simultaneously solve
the equation of the metric perturbation, which will complicate the
calculations. However, noting that the field perturbation is
related to the comoving curvature perturbation ${\cal R}$  \be
{H\over \dot{\phi}}\delta{\phi}= {\cal R}-\Phi,
\label{rp}\ee %in which ${\cal R}$ is the curvature perturbation.
thus we might firstly calculate ${\cal R}$ and then apply
Eq.(\ref{rp}) to obtain $\delta\phi$.
%which is
%related to $\Phi$ and $\delta \phi$ by ${\cal R}=\Phi
%+{H\delta\varphi\over {\dot \phi}}$.
%The quadratic
%action of the curvature perturbation $\cal R$ is \be S_2\sim \int
%d\eta d^3x {a^2M_P^2\epsilon\over c_s^2}\left({{\cal
%R}^\prime}^2-{c_s^2}(\partial {\cal R})^2\right), \ee which is
%actually universal for single field, e.g. \cite{GM}, where
%definition of $\epsilon$ is ${d\over dt}({1/ H})$.
The equation of $\cal R$ in momentum space is \cite{Muk},\cite{KS}
\be u_k^{\prime\prime} +\left(k^2-{z^{\prime\prime}\over z}\right)
u_k = 0, \label{uk}\ee  after $u_k \equiv z{\cal R}_k$ is defined,
where $'$ is the derivative with respect to the conformal time
$\eta=\int dt/a$, $z\equiv a\sqrt{2M_P^2\epsilon}$ and $\epsilon
=-{\dot H}/H^2$.
%We have
%$c_s^2=1$ for canonical scalar field.

When $k^2\simeq z^{\prime\prime}/z$, the perturbation mode is
leaving the horizon. When $k^2\ll z^{\prime\prime}/z$, the
solution of $\cal R$ given by Eq.(\ref{uk}) is \ba {\cal R}_k &
\sim &
C\,\,\,\,\, is\,\,\,{{\rm constant}}\,\,\,{ {\rm mode}}\label{C}\\
&or &\, D\int {d\eta\over z^2}\,\,\,\,\, is\,\,\,{{\rm
decaying/growing}}\,\,\,{ {\rm mode}} , \label{D}\ea where the $D$
mode is decaying or growing is dependent on the behavior of $z$.

The contraction phase may be regarded as \be a\sim (t_B-t)^{n},
\label{a}\ee where $t<t_B$ is negative, and the parameter $n$ is
constant, which is set only for the convenience of discussion.
Thus during the contraction with $n>{1\over 3}$, the amplitude of
$\cal R$ will be dominated by the growing mode,
\begin{eqnarray} {\cal R}_k \sim \int{d\eta\over z^2}\sim \left(t_B-t\right)^{1-3n}. \label{zeta}\end{eqnarray}
%When $n=1/3$ in Eq.(\ref{a}), the
%evolution of $\cal R$ becomes \ba {\cal R}_k\sim
%\ln\left(\frac{t_B-t_{Ci}}{t_B-t}\right).\ea
In principle, ${\cal R}_k$ will increase up to the end of this
contracting phase.

The amplitude of growing mode before the bounce may inherited by
the constant mode after the bounce \cite{Cai:2007qw},
\cite{Allen:2004vz},\cite{Xue:2013bva}, which is actually the
requirement that the curvature perturbation continuously comes
through the bounce. While in the expanding phase, the constant
mode is dominated, thus ${\cal R}_k$ will be unchanged until the
beginning of next contraction.

We will regard the beginning time of contracting phase as the
beginning of a cycle, and in one single cycle the universe will
orderly experience the contraction, bounce and expansion. In the
$j^{th}-1$ cycle, after the bounce, we have ${\cal R}_k^{j-1}$,
which will be constant up to the beginning of the $j^{th}$ cycle.
During the contraction of the $j^{th}$ cycle, ${\cal R}_k^j$ will
continue to increase. Thus after the bounce of the $j^{th}$ cycle,
we have \ba {\cal R}_k^{j}(t^{j}) & \simeq &
\left(\frac{t_{B}^{j}-t^{j}}{t_{B}^{j}-t_{Ci}^{j}}\right)^{1-3n_{j}}{\cal
R}_k^{j-1}(t_{Ce}^{j-1}) \nonumber\\ & \sim &
\prod_{l=2}^j\left(\frac{t_{B}^{l}-t^{l}}{t_{B}^{l}-t_{Ci}^{l}}\right)^{1-3n_{l}}
{\cal R}_k^{1}(t_{Ce}^{1}), \label{zeta1}\ea where $t_{Ci}^j$ and
$t_{Ce}^j$ are the beginning time and the end time of contracting
phase in the $j^{th}$ cycle, respectively. Here, we only consider
the perturbation mode, which is still outside the horizon all
along after it leaves the horizon in the $1^{th}$ cycle, or see
the details in Ref.\cite{Zhang:2010bb}.

The metric perturbation satisfies $({a\Phi_k\over H})^.= a {\cal
R}_k \epsilon$, e.g.\cite{GM}, thus we have $a\Phi_k/H=\int^t a
{\cal R}_k \epsilon \,\,dt^\prime$. The perturbation $\delta\phi$
of $\phi$ may be calculated as \ba \delta \phi_k & = &
\sqrt{2M_P^2\epsilon} {\cal R}_k\left(1-{{H\over a}\int^t a {\cal
R}_k \epsilon \,\,dt^\prime\over {\cal R}_k}\right)\simeq
\sqrt{2M_P^2\epsilon}{\cal R}_k\nonumber\\ & \sim &
\left(t_B-t\right)^{1-3n}, \ea where we have ${H\over a}\int^t a
{\cal R}_k \epsilon \,\,dt^\prime\sim {\cal R}_k$ in light of
Eqs.(\ref{a}) and (\ref{zeta}),
%\be {H\over {\dot \varphi}}\delta\varphi_k = \left( 1-{\Phi_k\over
%{\cal R}_k}\right){{\cal R}_k} \sim {{\cal R}_k}. \ee
which implies that the perturbation $\delta\phi$ of field
%\be \delta\varphi_k
%\sim \left({{\dot \varphi}\over H}\right)\int{d\eta\over z^2} \sim
%\label{dphik}\ee
will increase synchronously with ${\cal R}_k$.
%and ${{\dot
%\varphi}/ H}$ is constant,
%even if $\phi$ is oscillating
%around its minimum, in this case ${{\dot \varphi}/ H}$ is imposed
%but with a modulation $\sim Sin(M_\phi(t_B-t))$.
Thus we have \ba
\phi_k^{j}(t^{j})&\simeq &
\left(\frac{t_{B}^{j}-t^{j}}{t_{B}^{j}-t_{Ci}^{j}}\right)^{1-3n_{j}}{\phi}_k^{j-1}(t_{Ce}^{j-1})\nonumber\\
& \sim &
\prod_{l=2}^j\left(\frac{t_{B}^{l}-t^{l}}{t_{B}^{l}-t_{Ci}^{l}}\right)^{1-3n_{l}}
{\phi}_k^{1}(t_{Ce}^{1}). \label{phi1}\ea Both Eqs.(\ref{zeta1})
and (\ref{phi1}) indicates that the perturbation will be multipled
through cycles, and thus the rate of the amplification is quite
rapid.

\begin{figure}[htbp]
\includegraphics[scale=2,width=7.0cm]{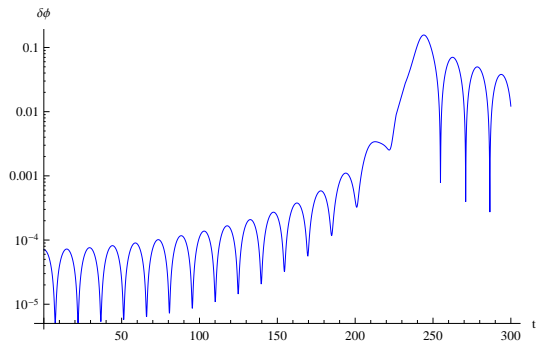}
\caption{The evolution of $\delta \phi$ with respect to the time
for the potential in the upper panel of Fig.\ref{fig:v}. The field
goes through one single AdS cycle, and its perturbation increases
during the corresponding contracting phase. }\label{fig:tv3}
\end{figure}

\begin{figure}[htbp]
\includegraphics[scale=2,width=7.0cm]{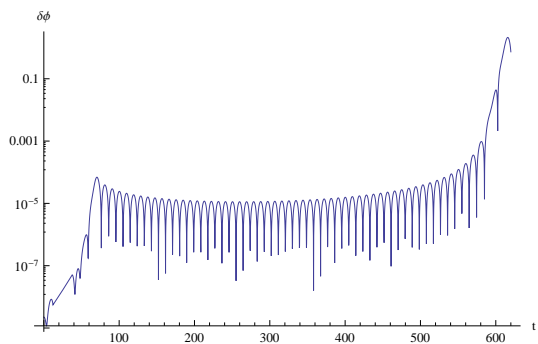}
\caption{The evolution of $\delta \phi$ with respect to the time
for the potential in the lower panel of Fig.\ref{fig:v}. The field
goes through two AdS cycles, and its perturbation increases during
the contracting phase of each AdS cycle. }\label{fig:tv4}
\end{figure}

We plot the evolutions of $\delta\phi$ in Figs.\ref{fig:tv3} and
\ref{fig:tv4} by numerically solving the set of the coupling
equations of $\Phi$ and $\delta\phi$, e.g.\cite{Liu:2012gu}. We
can clearly see that $\delta\phi$ is increasing during the
contraction and is almost constant during the expansion, thus the
net result is that the amplitude of perturbation is multipled
through cycles, which is consistent with Eq.(\ref{phi1}).

Here, we have assumed that the linear perturbation approximation
is satisfied all along. However, we will obviously see that it
will be broken in the $j^{th}_{Cutoff}$ cycles, which will set a
cutoff for the number of times of the AdS cycles of bubble
universe.

%\section{Proliferation of bubble universe }

%\subsection{The end of global cycle}
\subsection{Fragment of bubble universe}

We will estimate the value of $j_{Cutoff}$, in which cycle ${\cal
P}_{\cal R}\sim 1$ is arrived at.

During the contraction of AdS cycles, the universe will come
through the matter contracting phase, the kinetic phase and arrive
at the bounce. The perturbation amplitude will be amplified in the
matter contracting phase, in which \be {\cal R}_k\sim {1\over
t_B-t},\label{rk}\ee see Eq.(\ref{zeta}). The amplification of the
perturbation amplitude in kinetic phase is \be {\cal R}_k\sim
\int{d\eta\over z^2}\sim \ln(t_B-t), \ee which is negligible,
compared with that in matter contracting phase. In curvature
phase, since $a\sim 1/H$, the perturbation mode initially inside
the horizon will be still inside the horizon, which thus will not
be amplified. In addition, during the bounce the perturbation is
not also amplified \cite{Battarra:2014tga}.

\begin{figure}[htbp]
\includegraphics[scale=2,width=8.0cm]{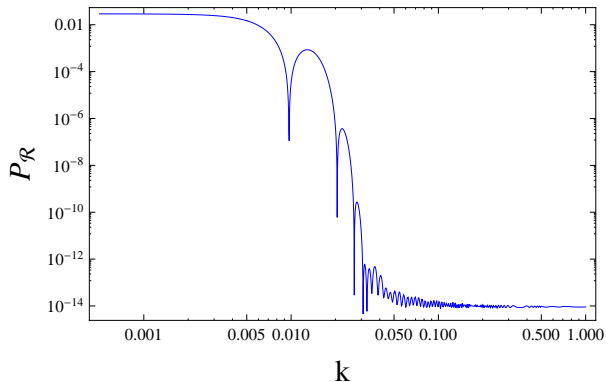}
\caption{ The snapshot of the power spectrum of curvature
perturbation in the $2^{th}$ AdS cycle with the evolution of the
background field in Fig.\ref{fig:phi}. }\label{fig:PR}
\end{figure}

When $k^2\gg z^{\prime\prime}/z$, the perturbation is deep inside
its horizon, $u_k$ oscillates with a constant amplitude. The
initial value is \be u_k\sim {1\over \sqrt{2k}} e^{-ik\eta}.
\label{ini}\ee When $k^2\ll {z^{\prime\prime}\over z}$, the
perturbation is far outside its horizon, the solution of
Eq.(\ref{uk}) can be obtained, which gives \be {\cal P}_{\cal
R}^{1/2}={k^{3/2}\over \sqrt{2\pi^2}}\left|{\cal R}_k\right|\sim
{H_{Kin}\over \sqrt{\epsilon_{Mat}}M_P},\label{PR}\ee where ${\cal
R}_k={u_k/ z}$ is used. The spectrum is scale invariant
\cite{Wands:1998yp},\cite{Finelli:2001sr}. This is the
perturbation spectrum after the bounce in the $1^{th}$ cycle. Thus
in the $j^{th}$ cycle, ${\cal P}_{\cal R}$ on the corresponding
scale is given by \be {\cal P}_{\cal R}^{1/2}={k^{3/2}\over
\sqrt{2\pi^2}}\left|{\cal R}^j_k\right|\sim
\left(\prod_{l\,\,=\,\,2}^{j}{H^l_{Kin}\over
H^l_{*}}\right){H^1_{Kin}\over \sqrt{\epsilon_{Mat}}M_P},
\label{multiP}\ee where Eq.(\ref{rk}) and $H\sim {1/(t_B-t)}$ are
used, and $H_*$ is the Hubble parameter at the beginning time of
the matter contracting phase. We may assume, for simplicity, that
$H_{Kin}$ in all cycles are equal, as well as $H_{*}$, which makes
Eq.(\ref{multiP}) becomes \be {\cal P}_{\cal R}^{1/2}\sim
\left({H_{Kin}\over H_{*}}\right)^j {H_{*}\over
\sqrt{\epsilon_{Mat}}M_P}. \ee Thus the breaking of the linear
perturbation approximation ${\cal P}_{\cal R}^{1/2}\gtrsim 1$
implies \be j_{Cutoff}\gtrsim { \ln^{-1}\left({H_{Kin}\over
H_{*}}\right)\ln{\sqrt{\epsilon_{Mat}}M_P\over H_{*}}}. \ee This
result implies that the larger the ratio between the scales that
the matter contracting phase begins and ends is, i.e. ${H_{Kin}/
H_{*}}$, the smaller $j_{Cutoff}$ is. When $j_{Cutoff}=1$, we have
\be H_{Kin}\sim M_P,\ee which is consistent with Eq.(\ref{PR}).
Thus unless the bounce occurs in Planck scale, it is hardly
possible that in the $1^{th}$ cycle the amplitude of curvature
perturbation increases up to 1. When \be {H_{Kin}\sim
H_{*}},\label{HH}\ee we may have $j_{Cutoff}\gg 1$, which,
however, is still a finite number. Here, (\ref{HH}) is equivalent
with the condition that the ratio of the maximal value of $a$ to
its minimal one is $ {\cal O}(1)$ in Ref.\cite{Graham:2011nb}. In
certain sense, the increase of the perturbation amplitude means
that an infinite cycles is impossible.

When $j_{Cutoff}=2$, we have \be H_*\lesssim \left({H_{Kin}\over
M_P}\right) H_{Kin}. \label{cut2}\ee Thus $H_*$ has to be large,
or ${\cal P}_{\cal R}^{1/2}\sim 1$ will be arrived in this cycle.
When ${H_{Kin}\over M_P}\sim 10^{-5}$, which is often required by
the bouncing model in which the observable universe may appear
after one single bounce, we have $H_*\lesssim 10^{-5} H_{Kin}$,
which seems easily satisfied. Thus it seems highly possible that
${\cal P}_{\cal R}^{1/2}\sim 1$ will occur at $j=2$.

We could reestimate this observation in light of the details of
the landscape. Here, $H_{Kin}$ is defined when ${\dot \phi}^2$
begins to dominate, i.e. ${{\dot \phi}^2} \sim {V_{Bar}}$, which
gives $H^2_{Kin}\sim {V_{Bar}\over M_P^2}$, in which $V_{Bar}$ is
the barrier separating different minima of effective potential.
While when the matter contracting phase begins, we have
$\rho_{Mat}\sim\rho_{Cur}\gtrsim {|{\Lambda}_*|}$, which gives $
H^2_{*}\gtrsim {|{\Lambda}_*|\over M_P^2}$. Thus Eq.(\ref{multiP})
becomes \ba {\cal P}_{\cal R}^{1/2} & \lesssim &
\left(\prod_{l\,\,=\,\,2}^{j}\sqrt{V^l_{Bar}\over |{
\Lambda}^l_*|}\right){\sqrt{V^l_{Bar}}\over
\sqrt{\epsilon_{Mat}}M_P}\nonumber\\ & = & \left({V_{Bar}\over |{
\Lambda}_*|}\right)^{j/2}{\sqrt{ \Lambda}_*\over
\sqrt{\epsilon_{Mat}}M_P}, \label{multiP1}\ea where in the second
line we assume that $V_{Bar}$ in all cycles are equal, as well as
${ \Lambda}_*$. When $j_{Cutoff}=2$, we have \be |{\Lambda}_*| <
\left({V_{Bar}\over M_P^4}\right) V_{Bar}, \label{vv}\ee where
$|{\Lambda}_*|$ is the depth of AdS minimum in a landscape, see
also the effective potential (\ref{V}). We see that if
${V_{Bar}^{1/4}\over M_P}\sim 10^{-3}$,
$|{\Lambda}_*|^{1/4}>10^{-3}V_{Bar}^{1/4} $ has to be required for
avoiding ${\cal P}_{\cal R}^{1/2}\sim 1$ in this cycle. Thus we
may conclude that for a high potential barrier $V_{Bar}$, unless
AdS minimum is far deep, generally we have $j_{Cutoff}=2$, i.e.
${\cal P}_{\cal R}^{1/2}\sim 1$ will be rapidly arrived within two
cycles.

%The above result is actually only a conservative estimate, since
%we have neglected the increasing of the perturbation amplitude in
%other phases. However, it is obvious that including these
%increases will consolidate the observation of $j_{Cutoff}=2$.

We numerically show the change of the power spectrum ${\cal
P}_{\cal R}$ through AdS cycles in Fig.\ref{fig:PR} with the
evolution of the background field in Fig.\ref{fig:phi}. The power
spectrum ${\cal P}_{\cal R}^{1}$ after the bounce in the $1^{th}$
cycle is scale invariant, which is given by a long period of the
matter contraction. During the contraction of the $2^{th}$ cycle,
the shape of spectrum on large scale is unchanged, i.e. still
scale invariant, but its amplitude is amplified. The spectrum on
small scale is also scale invariant, since the corresponding modes
are newly generated in the $2^{th}$ cycle. The spectrum of the
modes on middle scale will redshift. The result is consistent with
Eq.(\ref{multiP}).

\begin{figure}[htbp]
\includegraphics[scale=2,width=7.0cm]{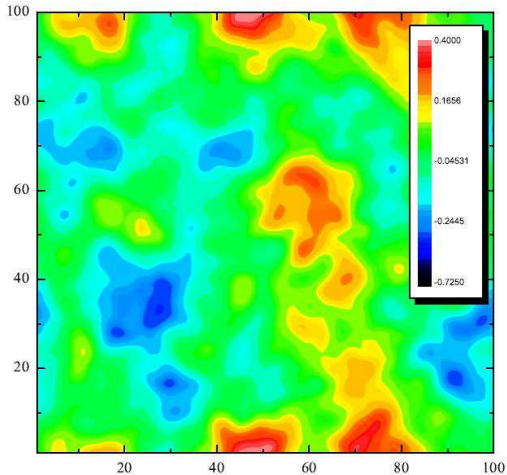}
\caption{ ${\cal R}(\vec{x})$ in position space, which reflects
the inhomogeneity of background when ${\cal P}_{\cal R}\sim 1$ on
large scale.}\label{fig:g4}
\end{figure}

When ${\cal P}_{\cal R}^{1/2}\sim 1$, it is hardly possible that
the background inside the bubble universe is still homogeneous.
The effect of the perturbation on background is plotted in
Fig.\ref{fig:g4} by transforming ${\cal R}_k$ into ${\cal
R}(\vec{x})$ in position space. We see that in the $2^{th}$ cycle,
the increase of the perturbation will eventually make the
initially homogeneous background become highly inhomogeneous, i.e.
fragment, thus the global cycle of the universe will inevitably
terminate.

Here, the existence of the matter contracting phase is crucial for
the amplification of perturbation. Thus based on the discussions
in II, we may conclude that the bubble universe going through AdS
cycles will become highly inhomogeneous, or fragment, within two
or three cycles, dependent on the details of landscape.

\subsection{New ``bubble" universes after fragment }

%\subsection{AdS cycles of bubble universe}

%\subsection{Proliferation of bubble universe }

We have showed that the bubble universe going through AdS cycles
will fragment at certain time $t_{Frag}$ within the $2^{th}$ or
$3^{th}$. We will see what is the resulting scenario.

The average square of the amplitude of field fluctuations at
$t_{Frag}$ is \ba & & <\delta\varphi^2_k>  =  {1\over
(2\pi)^3}\int \left|\delta\varphi_k\right|^2 d^3k \nonumber\\ &
\simeq & {1\over (2\pi)^3} \int^{aH}_{aH/e}
%{{\dot \varphi}^2\over H^2}
3M_P^2\left(1-{{H\over a}\int^t a {\cal R}_k \epsilon_{Mat}
\,\,dt^\prime\over
{\cal R}_k}\right)^2 \left|{\cal R}_k\right|^2 d^3k \nonumber\\
%\left({{\dot \phi}\over H}\right)^2{\cal P}_{\cal
%R} d\ln{k}
& = & 3 M_P^2 \,/4. \ea %Thus we have \be
%\sqrt{<\delta\varphi^2_k>}\simeq {{\dot \phi}\over H}\sim M_P, \ee
where $2\epsilon_{Mat}M_P^2=3M_P^2$ for the matter contraction and
${\cal P}_{\cal R}\sim 1$ are used. Thus at this moment it is
inevitable that the fields in different causal regions with length
$1/H_{Frag}$ will randomly jumps, in which $H_{Frag}={1\over
t_B-t_{Frag}}$ is the Hubble parameter at $t_{Frag}$, which
generally satisfies $H_*\ll H_{Frag}<H_{Kin}$.

%In principle, the eternal inflation will occur in any region of
%space where the amplitude of the density perturbation $\sim 1$.

%it seems that we have not $\Delta_H\phi\sim {{\dot \phi}/H}$,
%though ${{\dot \phi}/H}$ is constant. However,

When $t=t_{Frag}$, we have approximately \be {\cal P}_{\cal
R}^{1/2}\sim {H_{Frag}\over H_{*}}{H_{Kin}\over
\sqrt{\epsilon_{Mat}}M_P}\sim 1. \label{Frag}\ee When the matter
contraction begins, i.e. $t=t_*$, the length $l_*$ of the
homogeneous region inside the bubble should at least satisfy
$l_*\gtrsim 1/H_*$. Thus noting $l\sim a\sim (t_B-t)^{2/3}$ and
$H\sim 1/(t_B-t)$,
%noting that the curvature phase
%dose not affect this assumption, since $a\sim 1/H$.
at $t_{Frag}$ we have \be \left({l_{Frag}\over
1/H_{Frag}}\right)^3\simeq {H_{Frag}\over H_{*}}\sim
{\sqrt{\epsilon_{Mat}}M_P\over H_{Kin} } \gg 1, \ee where
Eq.(\ref{Frag}) is used. Thus at $t_{Frag}$, initial homogeneous
region will include lots of local regions with length
$1/H_{Frag}$.
%the different regions inside the bubble universe will evolve
%synchronously.

It is conceivable that in different local regions with the radius
$l_{Local}>1/H_{Frag}$, the field will jump to a different place
of its effective potential. In some regions, the field jumps to
certain place of its effective potential, which leads to
$\rho_{Local}>\rho_{Frag}$, thus \be {1/ H_{Local}} < {1/
H_{Frag}}<l_{Local}, \ee in which $1/H_{Local}$ is the Hubble
length of the corresponding region. This implies that the trapped
surface has got formed inside these regions. In this sense, such a
region actually corresponds to a new ``bubble" universe and will
continue to its contracting phase therein. While in other regions
we might have $\rho_{Local}<\rho_{Frag}$, i.e. $1/H_{Local}>
1/H_{Frag}$ thus initially there is not the trapped surface.
However, since the correspond region is contracting, $1/H_{Local}$
will shrink and ultimately become same order with $ 1/H_{Frag}$,
at this time the trapped surfaces also can get formed.

%and a new ``bubble" universe gets created.

Thus the initial bubble universe will fragment into a number of
local regions separated by domain walls, each of which actually
corresponds to a new ``bubble" universe. We may visually call this
as the proliferation of bubble universe. These ``bubble" universes
after proliferation will continue to go through cycles and then
fragment into newer ``bubble" universes until the field in
corresponding bubble lands at certain dS minimum.

\begin{figure}[htbp]
\includegraphics[scale=2,width=8.0cm]{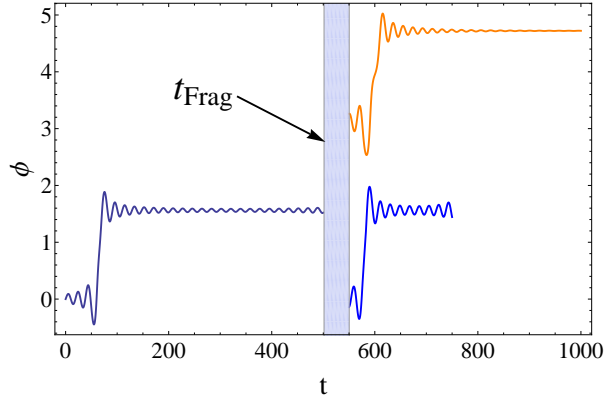}
\caption{The possible evolutions of fields inside different
``bubble" universes after the fragment, based on Fig.\ref{fig:phi}
for the effective potential in the lower panel of Fig.\ref{fig:v}.
The initial conditions of evolutions for them are that at
$t=t_{Frag}$ the value of field is shifted
$\sqrt{<\delta\varphi^2_k>}\sim 1\,\phi_{Unit}$ but $\dot\phi$ and
the sign of $\dot a$ are not altered. In Fig.\ref{fig:phi} the
field will go through two AdS cycles and eventually land at the
$\phi\simeq 3\phi_{Unit}$. However, if the effect of the field
fluctuation is considered, the case will be altered, the field
with $+ \phi_{Unit}$ shift will eventually land at the $\phi\simeq
5\phi_{Unit}$, while the field with $- \phi_{Unit}$ shift will
eventually land at the $\phi\simeq 1.5\phi_{Unit}$. Here, $t$ is
the global cosmic time, however, after $t_{Frag}$, in principle
each local region, or universe, has itself clock.
}\label{fig:c2power}
\end{figure}

We plot the possible evolutions of fields inside the new ``bubble"
universes in Fig.\ref{fig:c2power}, based on Fig.\ref{fig:phi}
with the assumption that at $t=t_{Frag}$ the value of field is
shifted $\sqrt{<\delta\varphi^2_k>}$, but $\dot\phi$ and the sign
of $\dot a$ are not changed. Here, $t$ is the global cosmic time,
however, after $t_{Frag}$, in principle each universe has itself
clock.

In Fig.\ref{fig:phi}, the effect of the field fluctuation is not
considered, the field will go through two AdS cycles and
eventually lands at the dS minimum. However, in
Fig.\ref{fig:c2power}, the case is altered, the global universe
will fragment within two AdS cycles, and after the fragment the
evolutions of fields inside different ``bubble" universes will be
different, after the bounce the field with the shift
$-\sqrt{<\delta\varphi^2_k>}$, or jumping back, will land at the
AdS minimum again, and the corresponding universe will begin next
cycle and might proliferate again, while the field with the shift
$+\sqrt{<\delta\varphi^2_k>}$, or jumping forward, will land at a
more distant part of the potential landscape, which might be a dS
minimum, as in Fig.\ref{fig:c2power}, and might not. Thus after
the proliferation the experiences of different ``bubble" universes
is not generally different, when some universes are in a phase of
the matter contraction, other universes might be in the phase of
the inflationary expansion or bounce.

It is also noted that the classical displacement of field during
one single cycle is approximately given by Eq.(\ref{De}), thus we
have
%\be
%\Delta\varphi=N_{Cycle}\,\Delta\varphi , \ee
\be {\sqrt{<\delta\varphi^2_k>}\over \Delta\varphi}= {{3\over
\sqrt{2}} \ln^{-1}\left({H_{B}^4\over H^2_{{Kin}}H^2_{{
Land}}}\right)}\sim 0.1-1. \ee This result indicates that when
${\cal P}_{\cal R}^{1/2}\sim 1$ the fluctuation of field will be
close to the order of its classical displacement during one single
cycle. In certain sense, this might provide an alternative insight
to the conclusion that the AdS bubble universe will fragment
within two or three cycles.

%why the fluctuation of field is effective to make the field jumped
%to adjacent minima.

\section{Discussion}

%\section{Implication for the eternal inflation scenario}

We have showed that in eternally inflating background, due to the
amplification of field fluctuation, the bubble universe going
through AdS cycles will inevitably fragment within two cycles or
three cycles and a number of new ``bubble" universes with
different vacua will come into being from these fragments. This
proliferation helps the eternal inflation to more rapidly populate
the whole landscape, in whichever corner of landscape it initially
happens.

%Though our numerical simulation is based on a ``toy" model of
%landscape, our result is model-independent, which has captured the
%basic ingredient of the full scenario.

%In the dS background, the nucleation probability of the new
%bubbles is expotentially low, which implies that after on single
%AdS cycle, if the bubble universe is in a dS state and inflating,
%we have to wait an expotentially long time to see a bubble with
%new vacuum. However, after a AdS cycle, if the bubble universe is
%still a AdS bubble, the case will be altered, the proliferation of
%bubble universe will occur, whose probability is not expotentially
%suppressed. We define this time as \be { \Delta t_{Frag}}={
%t_{Frag}-t_{*}}, \ee where $t_*$ is the time when the contracting
%phase begins in the $2^{th}$ cycle, Thus we have, from
%Eq.(\ref{Frag}), \be \Delta t_{Frag}\sim \left(1-{H_{Kin}\over
%\sqrt{\epsilon_{Mat}}M_P}\right){1\over H_*}\sim 1/H_*. \ee This
%implies that compared to the nucleation of bubbles in the dS
%background, the proliferation of bubble universe during AdS cycles
%is obviously more rapid. Thus

How to assign probabilities to different events in the eternally
inflating multiverse has been still a significant issue,
e.g.\cite{Freivogel:2011eg} for a review. The watcher measure
\cite{Garriga:2012bc}, in which all timelike geodesics are
required to extend to infinity, might be a promising avenue to
address the relevant problem.
%The AdS cycles in the eternally
%inflating spacetime
Our result solidifies the background of the watcher measure, in
which different regions of AdS bubble may ``transit" to different
vacua.

%It is generally thought that in the eternally inflating
%background, a phase of slow-roll inflation will occur inside some
%bubbles nucleated, which sets the initial condition of the ``big
%bang" evolution. However, our result brings an alternative route
%to the slow-roll inflation, in which the slow-roll inflation may
%be implemented inside new bubbles after the proliferation when the
%corresponding field lands at a plain of its effective potential.

It is conceivable that a phase of slow-roll inflation might occur
after the bounce. The bounce inflation may fit the observations
well, e.g.\cite{Piao:2003zm},\cite{Liu:2013kea}, which is
interesting for studying. We will back to this issue in details
elsewhere.

Recently, it has been argued \cite{Vilenkin:2014yva} that with AdS
bounce, the eternally inflating background is still
past-incomplete. However, with the proliferation of bubble
universe, it might be interesting to relook through this argument.

%we might live in an inflationary ``bubble" universe after the
%bounce.
%the proliferation. However, our simulation displays that the
%``bubble" is homogeneous but is initially highly anisotropic, thus
%a phase of slow-roll inflation should be required to make it both
%isotropic and flat. However, the residual anisotropy might be
%observable, e.g.\cite{Piao:2003zm},\cite{Liu:2013kea}, which might
%be a signature that we live in such a ``bubble" and is interesting
%for a detailed study.

\textbf{Acknowledgments}

This work is supported in part by NSFC under Grant No:11222546,
and in part by National Basic Research Program of China,
No:2010CB832804.

\section*{Appendix: A ``toy" model of AdS cycles}

%\subsection{Numerical simulation of AdS cycles}

\begin{figure}[htbp]
\includegraphics[scale=2,width=7.0cm]{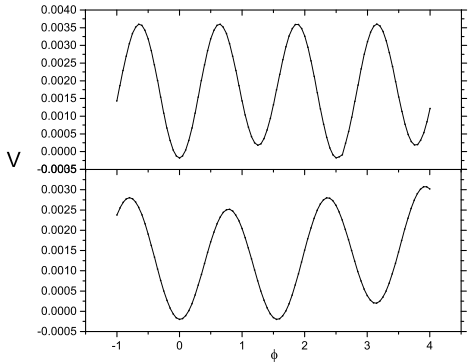}
\caption{In the upper panel, %we choose the parameter
%$\Lambda_*=1.8\times10^{-3}$, ${\tilde
%\Lambda}_*=1.8\times10^{-4}$, ${M_\phi\over \sqrt{\Lambda_*}}=5$,
we plot a potential, in which one dS minimum and one AdS minimum
alternate. In the lower panel, %we choose the parameter
%$\Lambda_*=7\times10^{-3}$, ${\tilde \Lambda}_*=1.4\times10^{-3}$,
%${M_\phi\over \sqrt{\Lambda_*}}=4$, thus
we plot a potential in which two AdS minima and two dS minima
alternate. Here and also through whole manuscript, the unit of
$\phi$ is $\phi_{Unit}=0.5M_P$ and the unit of $t$ is
$1/\phi_{Unit}$. } \label{fig:v}
\end{figure}

We will introduce a ``toy" model of AdS cycles, which will be used
to simulate the evolution of perturbation in Sec. III and IV. The
Lagrangian is \ba
\mathcal{L}=\frac{1}{2}\partial_{\mu}\phi\partial^{\mu}\phi-\left(\frac{1}{2}\partial_{\mu}\psi\partial^{\mu}\psi\right)^{2/3}-V(\phi).
\label{LL} \ea
%where the potential is only the function of the field $\phi$.
Here, we regard the potential as \ba
V=V_{Bar}\left(1-\cos({M_\phi\over
\sqrt{V_{Bar}}}\phi)\right)-{\Lambda}_*\cos({M_\phi\over
N_{Int}\sqrt{V_{Bar}}}\phi+\theta), \label{V}\ea which may be the
axion field in string theory,
e.g.\cite{Conlon:2006tq},\cite{Blumenhagen:2012kz}, where we
require that ${\Lambda}_*\ll V_{Bar}$ and $N_{Int}>0$ is the
integer. This potential has periodic minimal values, in which
$V_{Bar}$ sets the height of the potential barrier and ${
\Lambda}_*$ sets the depth of AdS minimum. When
$N_{Int}=2$, the potential % around the corresponding minima
around $\phi=0$ is approximately \be V\simeq
{M_{\phi}^2\phi^2\over 2}-{\Lambda}_*,\ee which is AdS-like, while
around its adjacent minimum, i.e. $\phi_1={2\pi
\sqrt{\Lambda_*}\over M_\phi}$, the potential is approximately \be
V\simeq {M_{\phi}^2(\phi-\phi_1)^2\over 2}+{\Lambda}_*,\ee which
is dS-like.
%is approximately \ba V & \simeq & {M_{\phi}^2\phi^2/
%2}-{\tilde\Lambda}_*, \,\,\,{for}\,\,\, \phi=0,
%\\ & & {M_{\phi}^2\phi^2/ 2}+{\tilde\Lambda}_*, \,\,\,{for}\,\,\, \phi=
%{2\pi \sqrt{\Lambda_*}\over M_\phi}. \ea
Thus in this potential the dS minimum and the AdS minimum
alternate, see the upper panel in Fig.\ref{fig:v}. While when
$N_{Int}=\,4$ and $\theta={\pi/\, 4}$, we have a potential in
which two AdS minima and two dS minima alternate, see the lower
panel in Fig.\ref{fig:v}.

The bounce is induced by the evolution of field $\psi$, which is
ghostlike. Here, we will regard $\psi$ as a purely classical field
to implement the background evolution of the nonsingular bounce
\cite{Cai:2007qw}, which is only significant around the bounce and
otherwise negligible.

However, it is generally thought that the appearance of such a
field is only the approximation of a fundamental theory below
certain physical cutoff. e.g. see G-bounce
\cite{Qiu:2011cy},\cite{Easson:2011zy},\cite{Cai:2012va},\cite{Osipov:2013ssa}
and super-bounce \cite{Koehn:2013upa}, and also
\cite{Rubakov:2014jja} for a review. The bounce is also
implemented in e.g.\cite{Gasperini:1992em} for Pre-big bang
scenario, \cite{Khoury:2001wf} for ekpyrotic scenario, and also
the string-inspired gravity
\cite{Biswas:2005qr},\cite{Piao:2003hh}, the multiscale gravity
\cite{Calcagni:2010bj}, other modified gravity
\cite{Bamba:2013fha}, see \cite{NB},\cite{Lehners:2011kr} for
reviews.

%However, for different bounce mechanisms, the scenario showed in
%text is not qualitatively altered, which is argued as follows.
The bounce generally occurs in a high energy scale, thus the
relevant physics are only reflected on the perturbation modes at
far small scale, while the perturbation modes which we care are
those at far large scales.
%In addition, the amplitude of
%perturbation is amplified cycle by cycle is determined not by the
%physics around the bounce, but by the behavior of the contracting
%phase.
Thus for different bounce mechanisms, the scenario showed
in text is not qualitatively altered.

When ${\dot\phi}^2$ is dominated, we have $\rho_\phi=C_{\phi}/a^6$
for $\phi$ field. While (\ref{LL}) implies
$\rho_{\psi}=C_{\psi}/a^{12}$. Thus the Friedmann equation is \be
3M_P^2\left({{\dot a}\over a}\right)^2 \simeq  {C_\phi\over
a^6}-{C_\psi\over a^{12}}.\label{F}\ee
%\be \int dt=\int
%\frac{da^6/6M_P\sqrt{3}}{\sqrt{c_{\phi}a^6-{c_\psi}}}.\ee
We may integrate it and have \be t-t_B={M_P\over \sqrt{3
C_{\phi}}}\sqrt{a^6-{C_\psi\over C_\phi}}.
%a(t)=\left(3M_P^2(t-t_B)^2C_\phi +{C_\psi\over
%C_\phi}\right)^{1/6},
\label{aa}\ee There is a bounce at $t=t_B$, $a^6_B={C_\psi\over
C_\phi}$. %When $a$ is far away from $a_B$, $a\sim (t_B-t)^{1/3}$,
%which is consistent with $\rho_\phi=C_{\phi}/a^6$.
When $\rho_B= {C_{\psi}\over C_\phi^2}$ is defined, Eq.(\ref{F})
can be rewrite as \be 3M_P^2\left({{\dot a}\over a}\right)^2
\simeq \rho_\phi-{\rho_\phi^2\over \rho_B}, \ee which is similar
to that in Ref.\cite{Shtanov:2002mb} and LQC
\cite{Ashtekar:2011ni}. Thus the evolution Eq.(\ref{aa}) of $a$ is
same with that in
Refs.\cite{Piao:2004me},\cite{Garriga:2013cix},\cite{Gupt:2013poa}.
When $\rho_\phi=\rho_B$, which is the bounce scale, the
contraction of universe halts and the expansion begins.

%and of the Coleman-Weinberg form \ba
%V_1=\frac{1}{4}\lambda\phi^4(ln\frac{|\phi|}{v}-\frac{1}{4})+\frac{1}{16}\lambda
%v^4 \ea see in Fig.\ref{fig:v2}

\begin{figure}[htbp]
\includegraphics[scale=2,width=7.0cm]{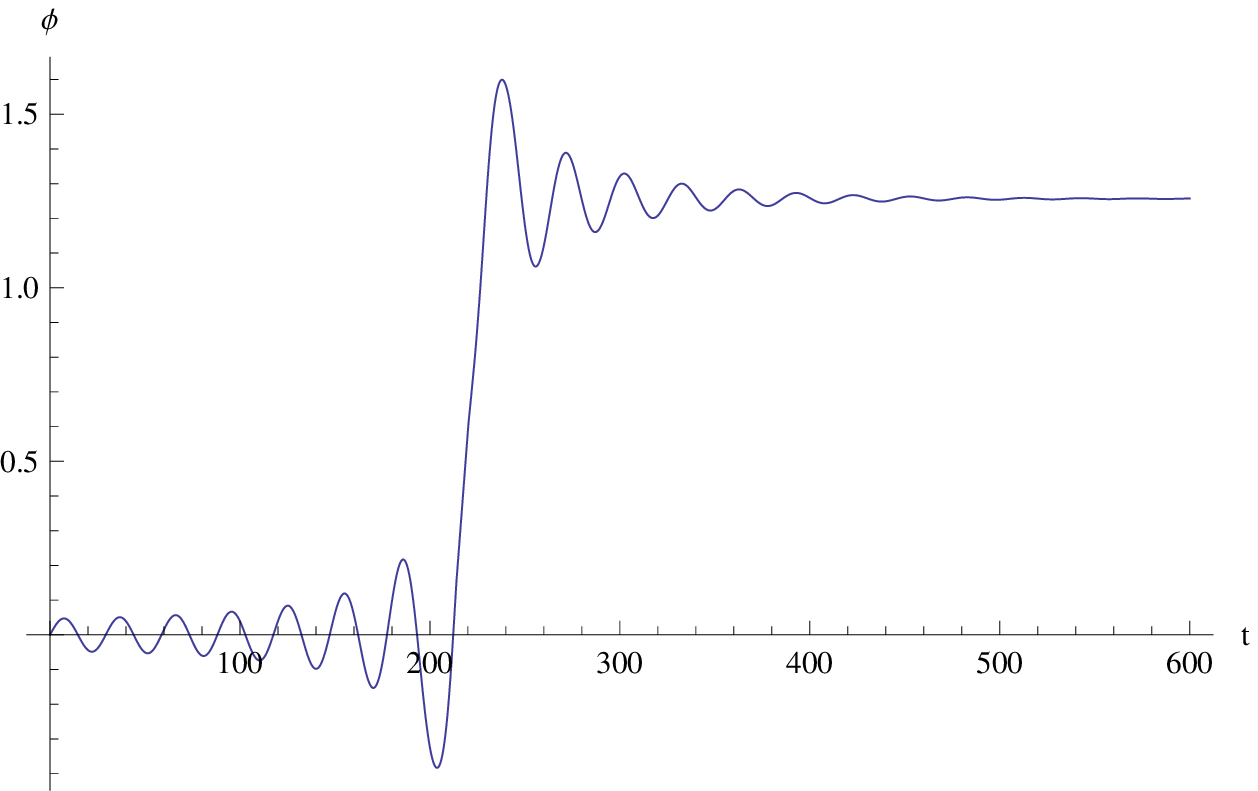}
\caption{The evolutions of $\phi$ with respect to the time for the
potential in the upper panel of Fig.\ref{fig:v}. The field, which
is initially in a AdS minimum of its effective potential, goes
through one single AdS cycle and finally lands at the dS minimum.
} \label{fig:pah}
\end{figure}

\begin{figure}[htbp]
\includegraphics[scale=2,width=7.0cm]{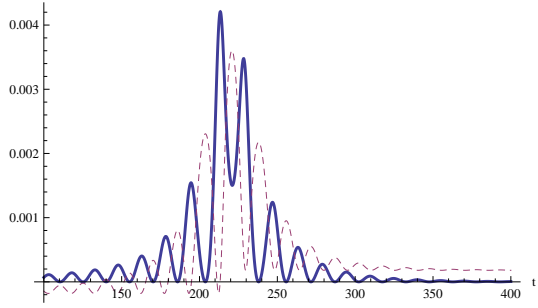}
\caption{The solid line is the evolution of kinetic energy for the
field in Fig.\ref{fig:pah}, while the dashed line is the evolution
of potential energy. }\label{fig:tv1}
\end{figure}

\begin{figure}[htbp]
\includegraphics[scale=2,width=7.0cm]{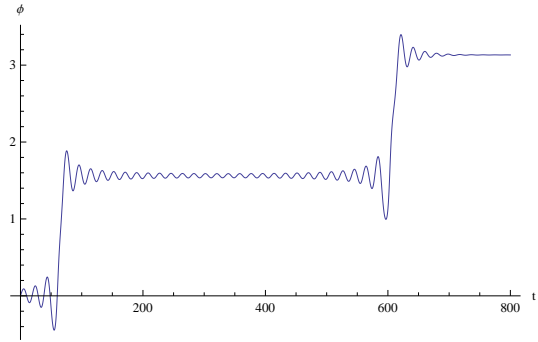}
\caption{The evolution of $\phi$ with respect to the time for the
potential in the lower panel of Fig.\ref{fig:v}. The field, which
is initially in a AdS minimum of its effective potential, goes
through two AdS cycles and finally lands at the dS
minimum.}\label{fig:phi}
\end{figure}

We plot the evolution of $\phi$ in Fig.\ref{fig:pah} for the
potential in the upper panel of Fig.\ref{fig:v}, as well as the
kinetic energy and the potential energy in Fig.\ref{fig:tv1}, and
the evolution of $\phi$ in Fig.\ref{fig:phi} for the potential in
the lower panel of Fig.\ref{fig:v}. During the contraction, the
field $\phi$ oscillates around a AdS minimum of its potential, we
have \be \phi \simeq {C_{Mat} \over a^{3/2}}Sin\left({M_\phi
(t_B-t)}\right), \label{osci}\ee where $C_{Mat}$ is the integral
constant. In light of Eq.(\ref{osci}), this oscillation lasts for
a period corresponds to $ M_{\phi}\left(t_{Kin}-t_*\right)\gg
\pi$, which equals to \be {M_{\phi}\over H_*}\gg 1 ,\ee since we
generally have $|t_*|\gg |t_{Kin}|$. Thus noting $H_*\simeq
\sqrt{|{\Lambda}_*|}/M_P$, in which ${\Lambda}_{*}$ is the depth
of AdS minimum, we have \be |{\Lambda}_*|\ll M_\phi^2 M_P^2
\,\,\lesssim V_{Bar}, \label{condition}\ee where the width of the
potential barrier is not larger than $M_P$ is required. The
condition (\ref{condition}) is consistent with Eq.(\ref{vv}),
which may be easily satisfied for any landscape, unless its AdS
minimum are far deep.

We see that during the AdS bounce, the field will ``fly" over the
potential barrier, and finally land at other place of its
effective potential. However, if where the field lands is a AdS
minimum again, it will ``fly" again until it finally lands at a dS
minimum. Thus we may have a AdS cycles, during which the bubble
universe goes through different AdS vacua.

%Here, the amplification of the perturbation amplitude during AdS
%cycles is crucial for the proliferation of bubble universe. In a
%contracting phase with $w\simeq 0$, the amplitude of the adiabatic
%perturbation will increase, which may be inherited by the constant
%mode in the expanding phase after the bounce. While in the
%landscape, the appearance of such a phase is inevitable, which is
%generally brought by the oscillating of the field around the AdS
%minimum of its effective potential.

\end{document}